\documentclass[prd,twocolumn,preprintnumbers,amsmath,amssymb,nofootinbib,superscriptaddress,notitlepage]{revtex4}
\usepackage[sort&compress]{natbib}
\usepackage{graphicx,color,dcolumn,booktabs,bm}
\usepackage{longtable,lscape}
\usepackage{txfonts}
\usepackage{overpic}
\usepackage{amssymb}
\usepackage{indentfirst}
\usepackage{feynmf}   
\usepackage{slashed}  
\usepackage{cases}
\usepackage{appendix}
\usepackage{color}
\usepackage{multirow}
\usepackage{longtable}
\usepackage{makecell}
\usepackage{ulem}
\usepackage{enumerate}
\usepackage{graphicx,color,dcolumn,booktabs,bm}
\usepackage[colorlinks,
            citecolor=blue,
            anchorcolor=red,
            menucolor=red,
            linkcolor=red,
            filecolor=red,
            runcolor=red,
            urlcolor=blue,
            frenchlinks=red]{hyperref}
\usepackage{epstopdf}
\usepackage{tikz}

\begin{document}

\title{Double-charm heptaquark states composed of two charmed mesons and one nucleon}

\author{Si-Qiang Luo}\email{luosq15@lzu.edu.cn}
\affiliation{School of Physical Science and Technology, Lanzhou University, Lanzhou 730000, China}
\affiliation{School of Mathematics and Statistics, Lanzhou University, Lanzhou 730000, China}
\affiliation{Research Center for Hadron and CSR Physics, Lanzhou University and Institute of Modern Physics of CAS, Lanzhou 730000, China}

\author{Li-Sheng Geng}\email{lisheng.geng@buaa.edu.cn}
\affiliation{School of Physics, Beihang University, Beijing 102206, China}
\affiliation{Beijing Key Laboratory of Advanced Nuclear Materials and Physics, Beihang University, Beijing, 102206, China}
\affiliation{School of Physics and Microelectronics, Zhengzhou University, Zhengzhou, Henan 450001, China}
\affiliation{Lanzhou Center for Theoretical Physics, Lanzhou University, Lanzhou 730000, China}

\author{Xiang Liu}\email{xiangliu@lzu.edu.cn}
\affiliation{School of Physical Science and Technology, Lanzhou University, Lanzhou 730000, China}
\affiliation{Lanzhou Center for Theoretical Physics, Lanzhou University, Lanzhou 730000, China}
\affiliation{Research Center for Hadron and CSR Physics, Lanzhou University and Institute of Modern Physics of CAS, Lanzhou 730000, China}
\affiliation{Key Laboratory of Theoretical Physics of Gansu Province, and Frontiers Science Center for Rare Isotopes, Lanzhou University, Lanzhou 730000, China}

\begin{abstract}
Inspired by the experimental discoveries of $T_{cc}$, $\Sigma_c(2800)$, and $\Lambda_c(2940)$ and the theoretical picture where they are $DD^*$, $DN$, and $D^*N$ molecular candidates, we investigate  the double-charm heptaquark system of $DD^*N$. We employ the one-boson-exchange model to deduce the pairwise $D$-$D^*$, $D$-$N$, and $D^*$-$N$ potentials and then study the $DD^*N$ system with the Gaussian expansion method. We find two good hadronic molecular candidates with   $I(J^P)=\frac{1}{2}(\frac{1}{2}^+)$ and $\frac{1}{2}(\frac{3}{2}^+)$ $DD^*N$ with only $S$-wave pairwise interactions. The conclusion remains unchanged even taking into account the $S$-$D$ mixing and coupled channel effects. In addition to providing the binding energies, we also calculate the root-mean-square radii of the $DD^*N$ system, which further support the molecular nature of the predicted states. They can be searched for at the upcoming LHC run 3 and run 4.
\end{abstract}

\maketitle

\section{Introduction}

The precision frontier of particle physics is always full of surprises. With increasing experimental precision, we have had first glimpses of exotic hadronic states.  {Remarkable progress in identifying new hidden-charm tetraquark states and observing a series of $XYZ$ charmoniumlike states, the $P_c$ pentaquark states~\cite{LHCb:2015yax,LHCb:2019kea}, the hexaquark candidates such as $Y(4630)$ observed in $e^+e^-\to \Lambda_c\bar{\Lambda}_c$ ~\cite{Belle:2008xmh} and $d^*(2380)$~\cite{WASA-at-COSY:2011bjg,WASA-at-COSY:2012uua,WASA-at-COSY:2013fzt,WASA-at-COSY:2014dmv,WASA-at-COSY:2014squ,WASA-at-COSY:2014qkg} has been made in the past two decades.} Obviously, the study of exotic hadronic states can provide valuable information on understanding the nonperturbative behavior of quantum chromodynamics. We refer interested readers to Refs.~\cite{Chen:2016qju,Liu:2013waa,Yuan:2018inv,Olsen:2017bmm,Guo:2017jvc,Hosaka:2016pey,Brambilla:2019esw,Chen:2022asf} for comprehensive reviews in this regard.

\begin{figure}[hptb]
\centering
\includegraphics[width=6.4cm]{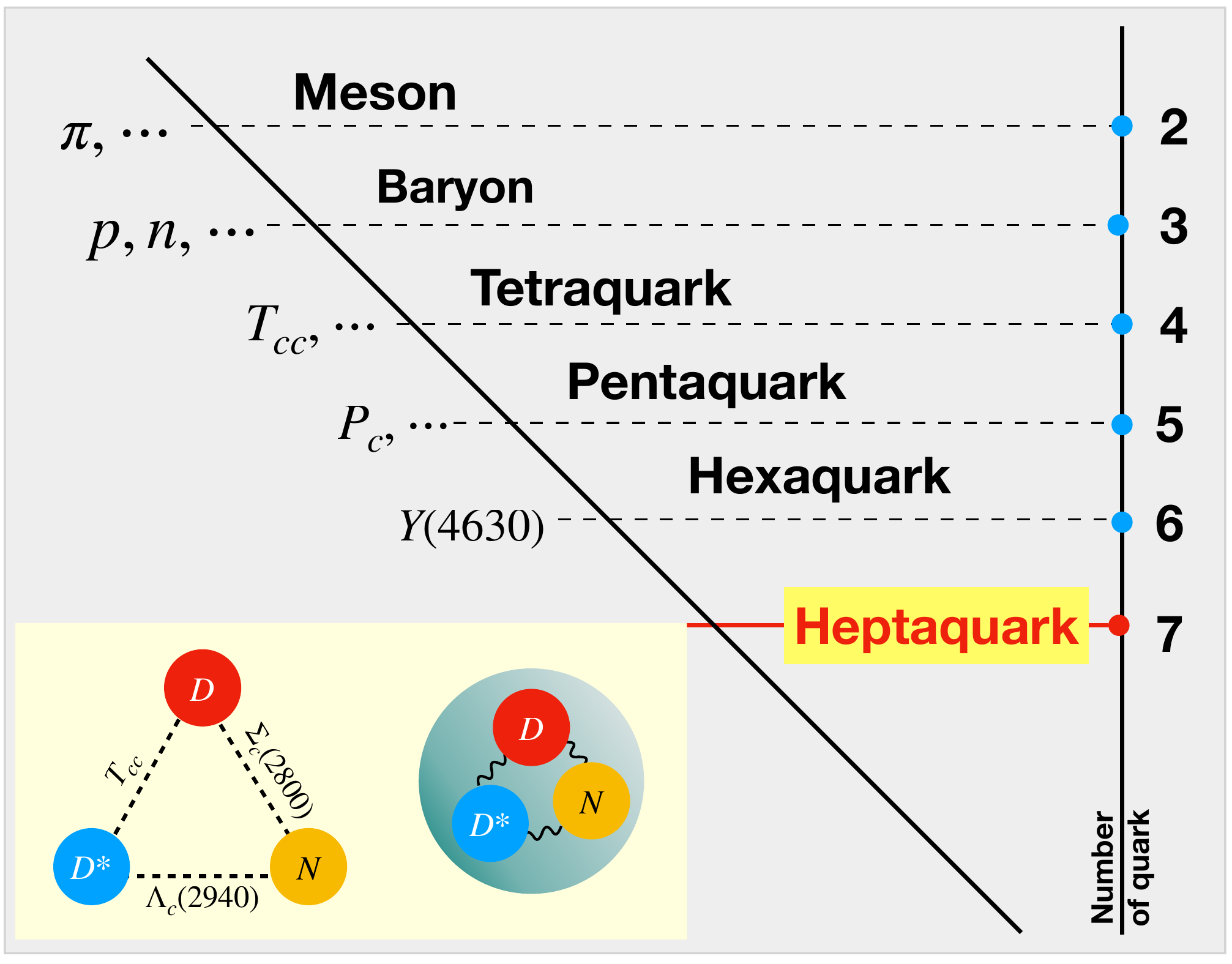}
\caption{Emergence of heptaquark states with the increasing number  of valence quarks.}
\label{fig1}
\end{figure}

Conventional mesons and baryons consist of two and three valence quarks. With the increasing number of valence quarks, we are now witnessing the emergence of  tetraquark, pentaquark, and heptaquark states. Facing the novel hadronic matter mentioned above  and given the exciting experimental observations, one wonders  whether or not there  exists exotic hadronic matter composed of more valence quarks, and heptaquark states naturally come to our mind (see Fig.~\ref{fig1}). Heptaquark states composed of seven quarks are not only a fantastic concept in few-body physics, but also a realistically allowed kind of exotic hadron matter in hadron physics, which deserve to be explored.

Scrutinizing all the newly discovered  hadrons, we notice three interesting ones, i.e., $T_{cc}^{+}$, $\Lambda_c(2940)$, and $\Sigma_c(2800)$. In 2021, the LHCb Collaboration observed the $T_{cc}^+$ state in the $D^0D^0\pi^+$ invariant mass spectrum \cite{LHCb:2021vvq,LHCb:2021auc}, which  is a good candidate for the $D\bar{D}^*$ molecular state~\cite{Li:2021zbw,Ren:2021dsi,Chen:2021vhg,Dai:2021vgf,Feijoo:2021ppq,Wu:2021kbu,Molina:2010tx,Li:2012ss,Liu:2019stu,Xu:2017tsr}. The discovery of the $\Lambda_c(2940)$ was reported by the BaBar Collaboration in the $D^0p$ invariant mass spectrum \cite{Aubert:2006sp} and confirmed in the $\Sigma_c(2455)\pi$ channel by the Belle Collaboration~\cite{Abe:2006rz} and in the $\Lambda_b^0\to D^0 p\pi^-$ process by the LHCb Collaboration~\cite{Aaij:2017vbw}. To understand its low low-mass,  the $\Lambda_c(2940)$ was suggested to be a $D^* N$ molecule~\cite{He:2006is,Dong:2010xv,He:2010zq,Ortega:2013fta,Wang:2020dhf} or a conventional baryon whose mass is lowered by the strong coupled-channel effect of $D^*N$ \cite{Luo:2019qkm}. The $\Sigma_c(2800)$ signal was first found by the Belle Collaboration in the $\Sigma_c^+\pi^{\pm,0}$ channels \cite{Belle:2004zjl}, which has also been interpreted as a $DN$ bound state~\cite{Wang:2018jaj,Zhang:2012jk,Dong:2010gu}. The above experimental and theoretical efforts have deepened our understanding of the relevant $D\bar{D}^*$, $D^*N$, and $DN$ interactions.

In this work we propose to study the double-charm heptaquark system of $D^*DN$  which is composed of two charmed mesons and one nucleon as shown in Fig. \ref{fig1}.  Intimately related to the observed $T_{cc}^+$, $\Lambda_c(2940)$, and $\Sigma_c(2800)$ states, the  open-charm heptaquark is an ideal platform to manifest how the change of the $DD^*$, $D^*N$, and $DN$ interactions affects the existence of an open-charm heptaquark state. The open-charm heptaquark  state is a typical few-body system, where the key issue is to deal with the
sub-two-body interactions of the  $D^*DN$ system. For this purpose, we adopt the Gaussian expansion method (GEM) \cite{Hiyama:2003cu,Hiyama:2012sma}, which is a powerful tool in studying few-body problems. Finally, the spectroscopy of the double-charm heptaquark states can be predicted.

This paper is organized as follows. After the Introduction, we  briefly introduce  the $D$-$D^{(*)}$ and $D^{(*)}$-$N$ potentials and the Gaussian expansion method in Sec.~\ref{formulism}. Then, in Sec.~\ref{numericalresults}, we present the binding energies and root-mean-square radii of the $DD^*N$ molecular candidates. We also study the sensitivity of the bound-state solutions to the coupling constants of the potentials in Sec.~\ref{solutionsconstants}. This paper ends with a short summary in Sec.~\ref{summary}.

\section{Formalism}\label{formulism}

To study the $DD^*N$ three-body problem, we should first  determine the $D{D}^*$, $DN$, and $D^*N$ interactions. Fortunately, the  observed $T_{cc}^+$, $\Lambda_c(2940)$, and $\Sigma_c(2800)$  have simulated studies of the $D{D}^*$ \cite{Li:2021zbw,Ren:2021dsi,Chen:2021vhg,Dai:2021vgf,Feijoo:2021ppq,Wu:2021kbu}, $DN$ \cite{Wang:2018jaj,Zhang:2012jk,Dong:2010gu}, and $D^*N$ ~\cite{He:2006is} interactions, respectively. In general, the one-boson-exchange (OBE) model provides a realistic framework where the effective interactions between hadron pairs can be deduced. First, we provide the following OBE effective potentials for the $D^{(*)}D^*$ system~\cite{Li:2012ss,Liu:2019stu,He:2010zq,Chen:2017jjn}:
\begin{equation}\label{VDmesonDmeson}
\begin{split}
V^{DD^*\to DD^*}=&-g_\sigma^2{\cal O}_{1}Y_\sigma+\frac{1}{2}\beta^2g_V^2{\cal O}_{1}{\cal H}_{V}\\
V^{DD^*\to D^*D}=&\frac{g^2}{3f_\pi^2}({\cal O}_{2}\hat{\cal O}+{\cal O}_{3}\hat{\cal P}){\cal H}_{P1}^{\prime}\\
&+\frac{2}{3}\lambda^2g_V^2(2{\cal O}_{2}\hat{\cal O}-{\cal O}_{3}\hat{\cal P}){\cal H}_{V1}^{\prime},\\
V^{DD^*\to D^*D^*}=&\frac{g^2}{3f_\pi^2}({\cal O}_{4}\hat{\cal O}+{\cal O}_{5}\hat{\cal P}){\cal H}_{P2}\\
&+\frac{2}{3}\lambda^2g_V^2(2{\cal O}_{4}\hat{\cal O}-{\cal O}_{5}\hat{\cal P}){\cal H}_{V2},\\
\end{split}
\end{equation}
For the interactions of $D^{(*)}$-$N$, we have
\begin{widetext}
\begin{equation}\label{VDmesonN}
\begin{split}
V^{DN\to DN}=&g_{\sigma NN}g_\sigma\left(1-\frac{{\bm \sigma}\cdot {\bf L}}{4m_N^2}\hat{\cal Q}\right)Y_\sigma+\left[-h_{vNN}\beta g_V-\frac{(h_{vNN}+2f_{vNN})\beta g_V}{8m_N^2}(\hat{\cal P}+2{\bm \sigma}\cdot{\bf L}\hat{Q})\right]{\cal H}_V^{\prime\prime},\\
V^{DN\to D^*N}=&\frac{g_{\pi NN}g_\pi}{3\sqrt{2}m_Nf_\pi}({\cal O}_{6}\hat{\cal O}+{\cal O}_{7}\hat{\cal P}){\cal H}_{P3}^{\prime\prime}+\left[-\frac{2h_{vNN}\lambda g_V}{m_N}{\cal O}_{8}\hat{\cal Q}+\frac{(h_{vNN}+f_{vNN})\lambda g_V}{3m_N}(2{\cal O}_{6}\hat{\cal O}-{\cal O}_{7}\hat{\cal P})\right]{\cal H}_{V3}^{\prime\prime},\\
V_{D^*N\to D^*N}=&g_{\sigma NN}g_\sigma{\cal O}_{9}\left(1-\frac{{\bm \sigma}\cdot {\bf L}}{4m_N^2}\hat{\cal Q}\right)Y_\sigma+\left[-h_{vNN}\beta g_V {\cal O}_{9}-\frac{(h_{vNN}+2f_{vNN})\beta g_V}{8m_N^2}{\cal O}_{9}(\hat{\cal P}+2{\bm \sigma}\cdot{\bf L}\hat{Q})\right]{\cal H}_V^{\prime\prime}\\
&+\frac{g_{\pi NN}g_\pi}{3\sqrt{2}m_Nf_\pi}({\cal O}_{10}\hat{\cal O}+{\cal O}_{11}\hat{\cal P}){\cal H}_{P}^{\prime\prime}+\left[-\frac{2h_{vNN}\lambda g_V}{3m_N}{\cal O}_{12}\hat{\cal Q}+\frac{(h_{vNN}+f_{vNN})\lambda g_V}{3m_N}(2{\cal O}_{10}\hat{\cal O}-{\cal O}_{11}\hat{\cal P})\right]{\cal H}_{V}^{\prime\prime}.
\end{split}
\end{equation}
\end{widetext}
In Eqs.~(\ref{VDmesonDmeson}) and (\ref{VDmesonN}), the ${\cal O}_i$'s are spin-dependent operators, which read explicitly
\begin{equation}
\begin{split}
&{\cal O}_{1}={\bm\epsilon}_4^\dagger\cdot{\bm\epsilon}_2,\\
&{\cal O}_2={\bm\epsilon}_3^\dagger\cdot{\bm\epsilon}_2,~{\cal O}_3=S({\bf r},{\bm \epsilon}_3^\dagger,{\bm \epsilon}_2),\\
&{\cal O}_{4}={\bm\epsilon}_3^\dagger\cdot(i{\bm\epsilon}_4^\dagger\times{\bm\epsilon}_2),~{\cal O}_5=S({\bf r},{\bm\epsilon}_3^\dagger,i{\bm\epsilon}_4^\dagger\times{\bm\epsilon}_2),\\
&{\cal O}_{6}={\bm\epsilon}_3^\dagger\cdot{\bm\sigma},~{\cal O}_{7}=S({\bf r},{\bm\epsilon}_3^\dagger,{\bm\sigma}),~{\cal O}_{8}={\bm\epsilon}_3^\dagger\cdot{\bf L}\\
&{\cal O}_{9}={\bm\epsilon}_3^\dagger\cdot{\bm\epsilon}_1,~{\cal O}_{10}=i{\bm\epsilon}_3^\dagger\times{\bm\epsilon}_1\cdot{\bm\sigma},\\
&{\cal O}_{11}=S({\bf r},i{\bm\epsilon}_3^\dagger\times{\bm\epsilon}_1,{\bm\sigma}),~{\cal O}_{12}=i{\bm\epsilon}_3^\dagger\times{\bm\epsilon}_1\cdot{\bf L},
\end{split}
\end{equation}
where $S({\bf r},{\bf a},{\bf b})=3({\bf a}\cdot \hat{\bf r})({\bf b}\cdot \hat{\bf r})-{\bf a}\cdot{\bf b}$ and ${\bf L}$ are tensor and orbital angular momentum operators, respectively. The ${\bm \epsilon}_i$ ($i=1,2$) and ${\bm \epsilon}_i^\dagger$ ($i=3,4$) are initial and final polarization vectors of the $D^*$ mesons, respectively. The conjugated potentials of Eqs.~(\ref{VDmesonDmeson}) and (\ref{VDmesonN}) could be obtained by the following interchange of polarization vectors
\begin{equation}
{\bm \epsilon}_1\leftrightarrow{\bm \epsilon}_3^\dagger,~~~
{\bm \epsilon}_2\leftrightarrow{\bm \epsilon}_4^\dagger.
\end{equation}
The ${\cal H}_{Pi} ^{(\prime,\prime\prime)}$ and ${\cal H}_{Vi} ^{(\prime,\prime\prime)}$ are defined as
\begin{equation}\label{Hfunction}
\begin{split}
{\cal H}_{Pi} ^{(\prime,\prime\prime)}=&{\cal C}_1^{(\prime,\prime\prime)}(I)Y_{\pi i}+\frac{1}{3}{\cal C}_0^{(\prime,\prime\prime)}(I)Y_{\eta i},\\
{\cal H}_{Vi} ^{(\prime,\prime\prime)}=&{\cal C}_1^{(\prime,\prime\prime)}(I)Y_{\rho i}+{\cal C}_0^{(\prime,\prime\prime)}(I)Y_{\omega i},
\end{split}
\end{equation}
respectively. In Eq.~(\ref{Hfunction}), the function $Y_i$ is written as
\begin{equation}
Y_i=\frac{{\rm e}^{-m_{Ei}r}}{4\pi r}-\frac{{\rm e}^{-\Lambda_i r}}{4\pi r}-\frac{\Lambda_i^2{\rm e}^{-\Lambda_i r}}{8\pi\Lambda_i}+\frac{m_{Ei}^2{\rm e}^{-\Lambda_i r}}{8\pi\Lambda_i}
\end{equation}
with $\Lambda_i=\sqrt{\Lambda^2-q_i^2}$ and $m_{Ei}=\sqrt{m_E^2-q_i^2}$. The $\Lambda$, $m_E$, and $q_i$ are the cutoff of the monopole form factor ${\cal F}(q^2,m_E^2)=(\Lambda^2-m_E^2)/(\Lambda^2-q^2)$, the mass of the exchanged meson, and the energy component of the exchanged momentum. The values of $q_i$ ($i=1,2,3$) are taken as $q_1=m_{D^*}-m_D$, $q_2=(m_{D^*}^2-m_D^2)/(4m_{D^*})$ and $q_3=(m_{D^*}^2-m_D^2)/(2(m_{D^*}+m_N))$. The isospin factors ${\cal C}^{(\prime,\prime\prime)}(I)$ are
\begin{equation}
\begin{split}
&{\cal C}_1(0)=-\frac{3}{2},~{\cal C}_1^\prime(0)=+\frac{3}{2},~{\cal C}_1^{\prime\prime}(0)=+\frac{3}{2},\\
&{\cal C}_1(1)=+\frac{1}{2},~{\cal C}_1^\prime(1)=+\frac{1}{2},~{\cal C}_1^{\prime\prime}(1)=-\frac{1}{2},\\
&{\cal C}_0(0)=+\frac{1}{2},~{\cal C}_0^\prime(0)=-\frac{1}{2},~{\cal C}_0^{\prime\prime}(0)=+\frac{1}{2},\\
&{\cal C}_0(1)=+\frac{1}{2},~{\cal C}_0^\prime(1)=+\frac{1}{2},~{\cal C}_0^{\prime\prime}(1)=+\frac{1}{2}.
\end{split}
\end{equation}
The operators $\hat{\cal O}$, $\hat{\cal P}$, and $\hat{\cal Q}$ are defined by
\begin{equation}
\hat{\cal O}=\frac{1}{r^2}\frac{\partial}{\partial r}r^2\frac{\partial}{\partial r},~~~
\hat{\cal P}=r\frac{\partial}{\partial r}\frac{1}{r}\frac{\partial}{\partial r},~~~
\hat{\cal Q}=\frac{1}{r}\frac{\partial}{\partial r}.
\end{equation}
To evaluate the above potentials, we need the values of the coupling constants and the masses of the mesons, which are collected in Table~\ref{parameters}.

\begin{table}[htbp]
\centering
\renewcommand\arraystretch{1.20}
\caption{Values of the coupling constants~\cite{Liu:2019stu,He:2010zq} and meson masses~\cite{ParticleDataGroup:2020ssz}. The signs are determined by the quark model.}
\label{parameters}
\begin{tabular*}{86mm}{@{\extracolsep{\fill}}llcccc}
\toprule[1.00pt]
\toprule[1.00pt]
\multicolumn{2}{c}{coupling constants}&\multicolumn{2}{c}{meson masses}\\
\midrule[0.75pt]
$\frac{g}{f_\pi}$=4.545 GeV$^{-1}$ &$g_{\pi NN}$   =-13.07         &$m_\pi$   =0.140 GeV &$m_\eta$ =0.548 GeV\\
$g_\sigma$       =0.76      &$g_{\sigma NN}$=-8.46          &$m_\sigma$=0.600 GeV &$m_D$    =1.867 GeV\\
$\beta g_V$      =5.2       &$h_{VNN}$      =3.25          &$m_\rho$  =0.770 GeV &$m_{D^*}$=2.009 GeV\\
$\lambda g_V$    =3.133 GeV$^{-1}$ &$f_{VNN}$      =19.83         &$m_\omega$=0.780 GeV &$m_N$    =0.939 GeV\\
\bottomrule[1pt]
\bottomrule[1pt]
\end{tabular*}
\end{table}

\begin{figure}[htbp]
\includegraphics[width=8.6cm]{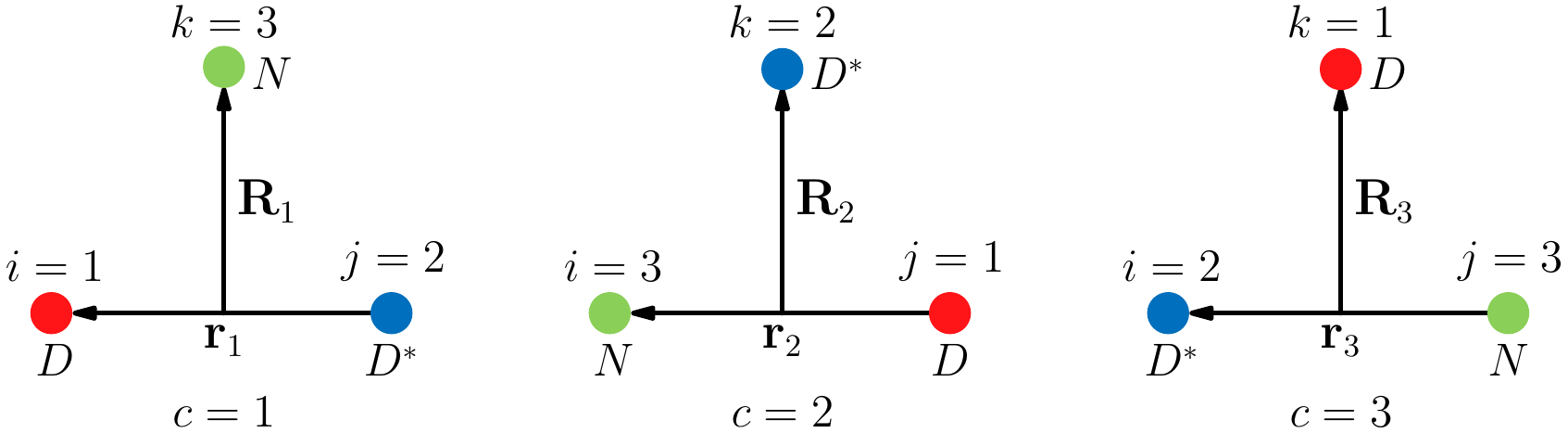}
\caption{Jacobi coordinates of the $DD^*N$ system.}
\label{Jacobi}
\end{figure}

To solve the three-body Schr\"odinger equation, we employ the Gaussian expansion method~\cite{Hiyama:2003cu,Hiyama:2012sma}. It is a popular method widely used in studying multibody hadronic molecular states~\cite{Wu:2019vsy,Wu:2020rdg,Wu:2020job,Wu:2021kbu,Wu:2021gyn,Wu:2021ljz,Wu:2021dwy,Luo:2021ggs}. The Jacobi coordinates of the $DD^*N$ system are presented in Fig.~\ref{Jacobi}.

The three-body Schr\"odinger equation reads
\begin{equation}\label{threebodySch}
\hat{H}\Psi_{JM}=E\Psi_{JM}
\end{equation}
with
\begin{equation}
\hat{H}=\hat{T}+V(r_1)+V(r_2)+V(r_3).
\end{equation}
$\Psi_{JM}$ is the total wave function, which is composed of three channels
\begin{equation}\label{PsiJM}
\Psi_{JM}=\sum_{c,\alpha}C_{c,\alpha} H_{t,T}^c\left[\chi^c_{s,S}\left[\phi_{nl}({\bf r}_c)\phi_{NL}({\bf R}_c)\right]_\lambda\right]_{JM},
\end{equation}
where the coefficient $C_{c,\alpha}$ is determined by the Rayleigh-Ritz variational method. The $c$ ($c=1,2,3$) represents the three channels in Fig.~\ref{Jacobi} and $\alpha=\{tT,sS,nN,lL\lambda\}$ is the quantum numbers of the basis. The $H_{t,T}^c$, $\chi^c_{s,S,M_S}$, and $\left[\phi_{nl}({\bf r}_c)\phi_{NL}({\bf R}_c)\right]_\lambda$ are flavor, spin, and spatial wave functions, respectively. The $\phi_{nlm_l}({\bf r}_c)$ and $\phi_{NLM_L}({\bf R}_c)$  read
\begin{equation}\label{gaussfun}
\begin{split}
\phi_{nlm_l}({\bf r}_c)=&N_{nl}r_c^le^{-\nu_nr_c^2}Y_{lm}(\hat{\bf r}_c),\\
\phi_{NLM_L}({\bf R}_c)=&N_{NL}R_c^Le^{-\lambda_NR_c^2}Y_{LM}(\hat{\bf R}_c),\\
\end{split}
\end{equation}
where $N_{nl}$ and $N_{NL}$ are normalization constants. In Eq.~(\ref{gaussfun}),  ${\bf r}_c$ and ${\bf R}_c$ are Jacobi coordinates, and  $\nu_n$ and $\lambda_N$  are Gaussian ranges, i.e.,
\begin{equation}
\begin{split}
\nu_n=&1/r_n^2,~r_n=r_1 a^{n-1}~(n=1,2\cdots n_{\rm max}),\\
\lambda_N=&1/R_N^2,~R_N=R_1 A^{N-1}~(N=1,2\cdots N_{\rm max}).
\end{split}
\end{equation}
With the above wave functions, all the Hamilton matrix elements could be expressed in simple forms. The details could be found in our previous work~\cite{Luo:2021ggs}.

\section{Numerical results}\label{numericalresults}

\begin{table*}
\caption{Bound state solutions of the $DD^*N$ system. The cutoff $\Lambda$, binding energies $E$, and root-mean-square radii are in units of GeV, MeV, and fm, respectively. The probabilities of $DD^*N$ and $D^*D^*N$ components are presented in the last two rows.}
\label{ResultsDDstarNTab}
\renewcommand\arraystretch{1.5}
\centering
\begin{tabular*}{172mm}{@{\extracolsep{\fill}}cc c ccccc ccccc cccc}
\toprule[1.00pt]
\toprule[1.00pt]
\multirow{2}{*}{$I$}&\multirow{2}{*}{$J^P$}
&\multicolumn{5}{c}{$S$-wave}
&\multicolumn{5}{c}{$S$-$D$ Mixing Effect}
&\multicolumn{4}{c}{Coupled Channel Effect}\\
\Xcline{ 3-7}{0.75pt}
\Xcline{8-12}{0.75pt}
\Xcline{13-16}{0.75pt}
&
&$\Lambda$&$E$&$r_{DD^*}$&$r_{DN}$&$r_{D^*N}$
&$\Lambda$&$E$&$r_{DD^*}$&$r_{DN}$&$r_{D^*N}$
&$\Lambda$&$E$&$P_{DD^*N}$ (\%)&$P_{D^*D^*N}$ (\%)
\\
\midrule[0.75pt]
\multirow{6}{*}{$\frac{1}{2}$}
&\multirow{3}{*}{$\frac{1}{2}^+$}
 &1.19&-0.11&4.31&10.24&10.20&1.02&-1.13&11.64&11.75&2.40&1.00&-0.41&97.75&2.25\\
&&1.24&-2.35&2.10&8.55&8.54&1.07&-21.19&11.56&11.59&1.26&1.05&-11.40&99.98&0.02\\
&&1.29&-6.62&1.38&6.34&6.33&1.12&-56.46&11.55&11.57&0.98&1.10&-40.43&$\sim$100&$\sim$0\\
\Xcline{2-16}{0.75pt}
&\multirow{3}{*}{$\frac{3}{2}^+$}
 &1.20&-0.37&3.81&10.24&10.18&0.89&-0.19&9.77&10.22&4.82&0.89&-0.22&99.98&0.02\\
&&1.25&-2.83&1.99&9.51&9.50&0.94&-3.25&9.22&9.34&2.47&0.94&-3.32&99.96&0.04\\
&&1.30&-7.12&1.37&9.13&9.12&0.99&-9.25&9.00&9.05&1.69&0.99&-9.44&99.83&0.17\\
\midrule[0.75pt]
\multirow{6}{*}{$\frac{3}{2}$}
&\multirow{3}{*}{$\frac{1}{2}^+$}
 &1.84&-0.11&10.71&11.12&5.17&1.77&-0.18&9.86&10.25&4.61&1.77&-0.21&99.99&0.01\\
&&1.89&-0.77&10.43&10.67&3.77&1.82&-0.94&9.63&9.85&3.47&1.82&-0.98&99.98&0.02\\
&&1.94&-1.72&10.24&10.38&2.87&1.87&-1.99&9.45&9.59&2.70&1.87&-2.04&99.98&0.02\\
\Xcline{2-16}{0.75pt}
&\multirow{3}{*}{$\frac{3}{2}^+$}
 &2.56&-1.32&2.13&9.01&9.00&1.90&-0.18&14.00&14.21&3.66&1.90&-0.18&$\sim$100&$\sim$0\\
&&2.61&-17.07&0.68&8.59&8.59&1.95&-1.16&13.77&13.90&2.82&1.95&-1.16&$\sim$100&$\sim$0\\
&&2.66&-41.08&0.47&8.53&8.53&2.00&-2.44&13.65&13.73&2.26&2.00&-2.44&$\sim$100&$\sim$0\\
\bottomrule[1pt]
\bottomrule[1pt]
\end{tabular*}
\end{table*}

With the deduced potentials, one could search for  bound-state solutions with the three-body Schr\"odinger equation. Before showing the numerical results, we would like to emphasize the following points~\footnote{These are relevant not only to studies of two-body hadronic molecular candidates, but also to those of three-body hadronic molecules.}:
\begin{enumerate}
\item The kinetic energy operator is equivalent to a repulsive potential. In general, it is difficult to form a higher partial wave hadronic molecular state. But three-body system contains two spatial degree of freedom, and the $S$-$D$ mixing effects introduce more bases and may affect the numerical results. Thus, in this work, we first consider the $S$-wave-only scheme. Then, the $S$-$D$ mixing effects are included.
\item In the $S$-$D$ mixing scheme, the tensor and spin-orbit terms can contribute to the matrix elements, which should also be taken into account  for completeness.
\item In our study, the cutoff $\Lambda$ is a crucial parameter in determining the existences of bound-state solutions. In our previous works~\cite{Li:2012ss,Chen:2015loa,Chen:2019asm,Chen:2018pzd,Chen:2017jjn,Yasui:2009bz}, the  cutoff $\Lambda$ is suggested to be about 1 GeV, whose value is supported by the studies of typical hadronic molecular candidates, such as deuteron~\cite{Yasui:2009bz}, $P_c$~\cite{Chen:2015loa,Chen:2019asm}, and $T_{cc}$~\cite{Wu:2021kbu}. If a bound-state solution is obtained with $\Lambda\approx 1$ GeV, this state could be viewed as a good molecular candidate.

\item {In our study, the spatial wave functions of $D$, $D^*$, and $N$ are not considered. We employ the center-of-mass of each hadron as the position in the Jacobi coordinates as depicted in Fig.~\ref{Jacobi}. An ideal molecular candidate is that the two constituent hadrons should not overlap too much in the spatial distributions. Thus the root-mean-square radii which describe the sizes of the $D$, $D^*$, $N$, and the three-body molecular states are important parameters when finding out the bound state solutions. With the Godfrey-Isgur model \cite{Godfrey:1985xj} and Capstick-Isgur model \cite{Capstick:1986ter}, the root-mean-square radii of $D$, $D*$, and $N$ could be estimated to be $r_D\approx0.40$ fm, $r_{D^*}\approx0.46$ fm, and $r_N\approx 0.70$ fm, respectively. In this work, we not only present the binding energies, but also calculate the root-mean-square radii between each pair of the three constituents, i.e., $r_{DD^*}$, $r_{DN}$, and $r_{D^*N}$. For a good candidate of  molecular state, we expect that there exist the relations $r_{DD^*}\gtrsim r_D+r_{D^*}$, $r_{DN}\gtrsim r_D+r_{N}$, and $r_{D^*N}\gtrsim r_{D^*}+r_{N}$ associated with the bound state solutions.}
\end{enumerate}

The numerical results are presented in Table~\ref{ResultsDDstarNTab}. We note that it is easy to obtain a bound  $I(J^P)=\frac{1}{2}(\frac{1}{2}^+)$ $DD^*N$ state. According to Table~\ref{ResultsDDstarNTab}, we can deduce the following:
\begin{enumerate}
\item There exist bound-state solutions in the $S$-wave $I(J^P)=\frac{1}{2}(\frac{1}{2}^+)$ $DD^*N$ configuration for a  cutoff $\Lambda\approx 1.2$ GeV. The root-mean-square radii of $DD^*$, $DN$, and $D^*N$ are about 3.5, 9, and 9 fm, respectively. Since the cutoff is approximately 1 GeV and the root-mean-square radii are several femtometers, the $S$-wave $I(J^P)=\frac{1}{2}(\frac{1}{2}^+)$ $DD^*N$ bound state can be viewed as a good molecular candidate.
\item In the $S$-$D$ mixing and coupled-channel schemes, there exist as well bound-state solutions for a cutoff $\Lambda\approx 1$ GeV. We note that these effects increase the strength of the attractive potentials such that a smaller cutoff is needed to yield  the same binding energies as those of the $S$-wave-only scheme. Similar phenomena have been observed in Refs.~\cite{Wang:2021yld,Wang:2021aql,Yang:2021sue,Wang:2021hql}.
\end{enumerate}

\begin{figure}
\includegraphics[width=8.6cm]{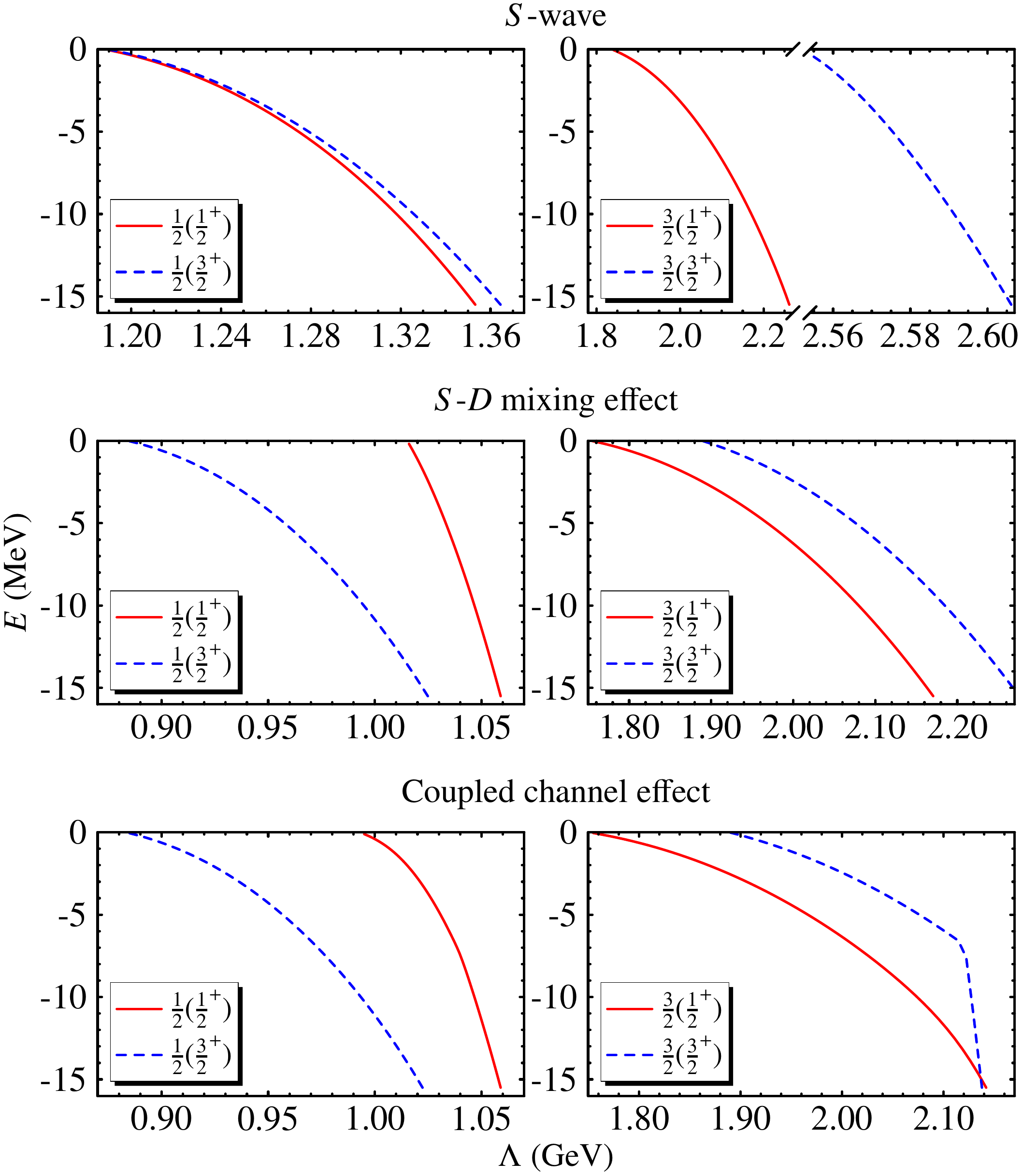}
\caption{Binding energies $E$ as functions of the cutoff $\Lambda$.}
\label{ResultsDDstarNFig}
\end{figure}

For the $S$-wave $I(J^P)=\frac{1}{2}(\frac{3}{2}^+)$ $DD^*N$ configuration, we find bound-state solutions when the cutoff $\Lambda$ is approximately 1.20 GeV. The root-mean-square radii of $DD^*$, $DN$, and $D^*N$ are several femtometers. Thus, the $S$-wave $I(J^P)=\frac{1}{2}(\frac{3}{2}^+)$ $DD^*N$  bound state is also a good molecular candidate. By decomposing the $S$-wave $I(J^P)=\frac{1}{2}(\frac{3}{2}^+)$ $DD^*N$ configuration, we find an important $D^*N$ substructure with $I(J^P)=0(\frac{3}{2}^+)$, which could be related to the hadronic molecular candidate $\Lambda_c(2940)$. According to the coupling constant determined in the quark model, the pion exchange interaction of $I(J^P)=0(\frac{3}{2}^+)$ $D^*N$ is attractive but repulsive for the $I=1$ configuration. Therefore, $\Lambda_c(2940)$ is good candidate for a  $I(J^P)=0(\frac{3}{2}^+)$ $D^*N$ molecular state.

When we consider the $S$-$D$ mixing effect, we find a loosely $DD^*N$ bound state  of $I(J^P)=\frac{1}{2}(\frac{3}{2}^+)$  for a $\Lambda\approx$ 0.89 GeV. But the coupled-channel effect only plays a minor role in this case, which is manifested by the small fractions of the $D^*D^*N$ component.

In our previous three-body studies of the $D^{(*)}D^{(*)}D^*$ systems,  we showed that the systems with higher isospins are much more difficult to bind than the systems of lower isospins, regardless of whether one considers only $S$-wave interactions, $S$-$D$ mixings, or coupled-channel effects. The same can be said about the $DD^*N$ system.

For the $S$-wave $I(J^P)=\frac{3}{2}(\frac{1}{2}^+)$ $DD^*N$ configuration, we obtain bound-state solutions for a cutoff $\Lambda\approx$1.85 GeV, which is larger than that in the scenarios of $I(J^P)=\frac{1}{2}(\frac{1}{2}^+)$, $\frac{1}{2}(\frac{3}{2}^+)$. The larger cutoff needed or the weaker interaction for isospin 3/2 can be attributed to the following two factors:
\begin{enumerate}[i.]
\item The pion exchange potentials of $1(1^+)$ $DD^*$ and $1(\frac{3}{2}^-)$ $D^*N$ are repulsive.
\item The attractiveness of the $I(J^P)=1(\frac{1}{2}^-)$ $DN$ configuration is also weaker than the $I=0$ configuration.
\end{enumerate}
When the $S$-$D$ mixing and coupled-channel effects are included,  the cutoff needed  to yield loosely bound-state solutions is decreased to  $\Lambda\approx1.77$ GeV

Among the four states, the $I(J^P)=\frac{3}{2}(\frac{3}{2}^-)$ $DD^*N$ configuration is the most difficult to form a bound state, which is reflected by the large cutoff $\Lambda$ in Table~\ref{ResultsDDstarNTab} and Fig.~\ref{ResultsDDstarNFig}. In the $S$-wave-only scheme, bound-state solutions start to emerge as the cutoff reaches  about 2.56 GeV, which is much larger than that of the $I=\frac{1}{2}$ configuration. When  the $S$-$D$ mixing and coupled-channel effects are taken into account, the value of the cutoff $\Lambda$ decreases to about 1.90 GeV.

{Searching for the predicted $DD^*N$ molecular candidates is a challenging issue.
We spell out the isospin wave functions for the convenience of  the following discussions. For the $DD^*N$ system with $I=\frac{1}{2}$, its isospin wave function  $|I_{DD^*},I,I_3\rangle$ reads as   
\begin{equation}\label{isospin12}
\begin{split}
\left|0,\frac{1}{2},+\frac{1}{2}\right\rangle=&\frac{1}{\sqrt{2}}D^0D^{*+}p-\frac{1}{\sqrt{2}}D^+D^{*0}p,\\
\left|0,\frac{1}{2},-\frac{1}{2}\right\rangle=&\frac{1}{\sqrt{2}}D^0D^{*+}n-\frac{1}{\sqrt{2}}D^+D^{*0}n,\\
\left|1,\frac{1}{2},+\frac{1}{2}\right\rangle=&\frac{1}{\sqrt{6}}D^0D^{*+}p+\frac{1}{\sqrt{6}}D^+D^{*0}p+\sqrt{\frac{2}{3}}D^+D^{*+}n,\\
\left|1,\frac{1}{2},-\frac{1}{2}\right\rangle=&\frac{1}{\sqrt{6}}D^0D^{*+}n+\frac{1}{\sqrt{6}}D^+D^{*0}n+\sqrt{\frac{2}{3}}D^0D^{*0}p.\\
\end{split}
\end{equation}
And the isospin wave functions of the $DD^*N$ system with $I=\frac{3}{2}$ are
\begin{equation}\label{isospin32}
\begin{split}
\left|1,\frac{3}{2},+\frac{3}{2}\right\rangle=&D^+D^{*+}p,\\
\left|1,\frac{3}{2},+\frac{1}{2}\right\rangle=&\frac{1}{\sqrt{3}}D^0D^{*+}p+\frac{1}{\sqrt{3}}D^+D^{*0}p-\frac{1}{\sqrt{3}}D^+D^{*+}n,\\
\left|1,\frac{3}{2},-\frac{1}{2}\right\rangle=&\frac{1}{\sqrt{3}}D^0D^{*0}p+\frac{1}{\sqrt{3}}D^0D^{*+}n-\frac{1}{\sqrt{3}}D^+D^{*0}n,\\
\left|1,\frac{3}{2},-\frac{3}{2}\right\rangle=&D^0D^{*0}n.\\
\end{split}
\end{equation}
By Eqs.~(\ref{isospin12})-(\ref{isospin32}), the $I=\frac{1}{2}$ and $I=\frac{3}{2}$ $DD^*N$ systems contain singly and doubly charged states, while the $I=\frac{3}{2}$ $DD^*N$ systems also contain a triply charged state and a neutral state. In this scheme, the triply charged and neutral channels are special when discussing decays of the $I=\frac{3}{2}$ $DD^*N$ states.

In general, the three-body molecular states have abundant decay channels. In this work, we mainly discuss those channels that are both kinematically and Okubo-Zweig-Iizuka allowed, which are summarized as follows:
\begin{enumerate}
\item If the $DD^*N$ molecular states have extremely shallow binding energies, they may decay into $T_{cc}^+p$. Then, the $T_{cc}^+$ could be observed in the $D^0D^0\pi^+$ or $D^0D^+\gamma$ final states. On the other hand, both the theoretical studies \cite{Ren:2021dsi,Li:2012ss,Liu:2019stu}
 and experimental analyses \cite{LHCb:2021auc} imply that the $T_{cc}^+$ has $I=0$. Thus, the total isospin of the $T_{cc}^+p$ channel is $1/2$. If the $DD^*N$ molecular state can be found in the $T_{cc}^+p$ channel, this $DD^*N$ molecular state must have $I=1/2$.
 
\item If the threshold of the $T_{cc}^+p$ channel is higher than the masses of the $DD^*N$ molecular states, the $T_{cc}^+p$ is kinematically forbidden. Since  the $D^*$ mass is about 140 MeV higher than that of the $D$ meson, the $DD^*N$ molecular states may decay into $DDp$, $DD\pi p$, and $DD\gamma p$.

\item In 2017, the doubly charmed baryon $\Xi_{cc}^{++}$ was observed by the LHCb Collaboration~\cite{LHCb:2017iph}. The thresholds of the $\Xi_{cc}^{++}\pi$ and $\Xi_{cc}^{++}\pi\pi$ channels are about 3760 and 3900 MeV, respectively, which are below the masses of the $DD^*N$ molecular states, which have  masses of about 4818 MeV. Obviously, studying the neutral $DD^*N$ molecular state, and single, double, and triple charged $DD^*N$ molecular states by these $\Xi_{cc}^{++}\pi^-\pi^-$, $\Xi_{cc}^{++}\pi^-$, $\Xi_{cc}^{++}\pi^+\pi^-$, and $\Xi_{cc}^{++}\pi^+$ decay channels, respectively, are possible.

\item For the discussed $DD^*N$ molecular system, the $D^{(*)+}p$ and $D^{(*)0}n$ components could annihilate into charmed baryons $\Sigma_c^{++}$ and $\Sigma_c^0$, respectively, while the $D^{(*)0}p$ component can couple with both $\Sigma_c^{+}$ and $\Lambda_c^+$. As a result, the $DD^*N$ molecular states may decay into a singly charmed baryon together with a $D$ meson. Because of the isospin conservation, the $I=\frac{1}{2}$ $DD^*N$ molecular states can decay into $\Lambda_c D^{(*)}$ and $\Sigma_c^{(*)} D^{(*)}$, while the $I=\frac{3}{2}$ states can only decay into $\Sigma_c^{(*)} D^{(*)}$. Considering that the $\Lambda_c D^{(*)}$ channels have $I=\frac{1}{2}$, the $\Lambda_c D^{(*)}$ channels are crucial to distinguish the isospins of the $DD^*N$ molecular states. As shown above, the $I=\frac{1}{2}$ systems are much more easier to form bound states than the $I=\frac{1}{2}$ systems. We suggest to search for the $DD^*N$ molecular states via the $\Lambda_c D^{(*)}$ channels. In addition, the $DD^*N$ molecular states with $I=\frac{3}{2}$ have typical decay channels $\Sigma_c^{++}D^{(*)+}$ and $\Sigma_c^0D^{(*)0}$.
\end{enumerate}
}

\section{Sensitivities of bound-state solutions to the coupling constants}\label{solutionsconstants}

In the previous section, we studied the dependence of binding energies on the cutoff $\Lambda$. However, the cutoff $\Lambda$ is not the only parameter in our study. The coupling constants are also crucial for the existences of bound states. There are eight coupling constants in our OBE potentials, i.e., $g/f_\pi$, $g_\sigma$, $\beta g_V$, $\lambda g_V$, $g_{\pi NN}$, $g_{\sigma NN}$, $h_{VNN}$, and $f_{VNN}$. With the experimental partial decay width of $D^*\to D\pi$, we obtain $g/f_\pi=4.545\;{\rm GeV}^{-1}$, which is close to the value $\frac{g}{f_\pi}=-\frac{3g_{\pi NN}}{5\sqrt{2}m_N}=5.905\;{\rm GeV}^{-1}$ determined in the quark model. In addition, the $\beta g_V=5.2$ used in this work is approximate to the quark model result $g_V=2h_{vNN}=6.50$. However, $g_\sigma=0.76$ and $\lambda g_V=3.13\;{\rm GeV}^{-1}$ are much less than the quark model results $g_\sigma=-\frac{1}{3}g_{\sigma NN}=2.82$ and $\lambda g_V=\frac{3(f_{VNN}+h_{VNN})}{10m_N}=7.37\;{\rm GeV}^{-1}$.

Since there are eight coupling constants, it is difficult to vary them simultaneously to study the impact on the three-body results. A more appropriate  approach is to vary one coupling constant while fixing the remaining and then  search for bound states. As discussed in the last paragraph, the values $g/f_\pi$ and $\beta g_V$ are consistent with the quark model predictions from $g_{\pi NN}$ and $h_{VNN}$, respectively. In this sense,  the values of $g/f_\pi$, $\beta g_V$, $g_{\pi NN}$, and $h_{VNN}$ are reasonable, whereas $g_\sigma$, $\lambda g_V$, $g_{\sigma NN}$, and $f_{VNN}$ need to be better understood. Thus we mainly discuss the sensitivities of bound-state solutions to $g_\sigma$, $\lambda g_V$, $g_{\sigma NN}$, and $f_{VNN}$.

\begin{figure}[t]
\includegraphics[width=8.6cm]{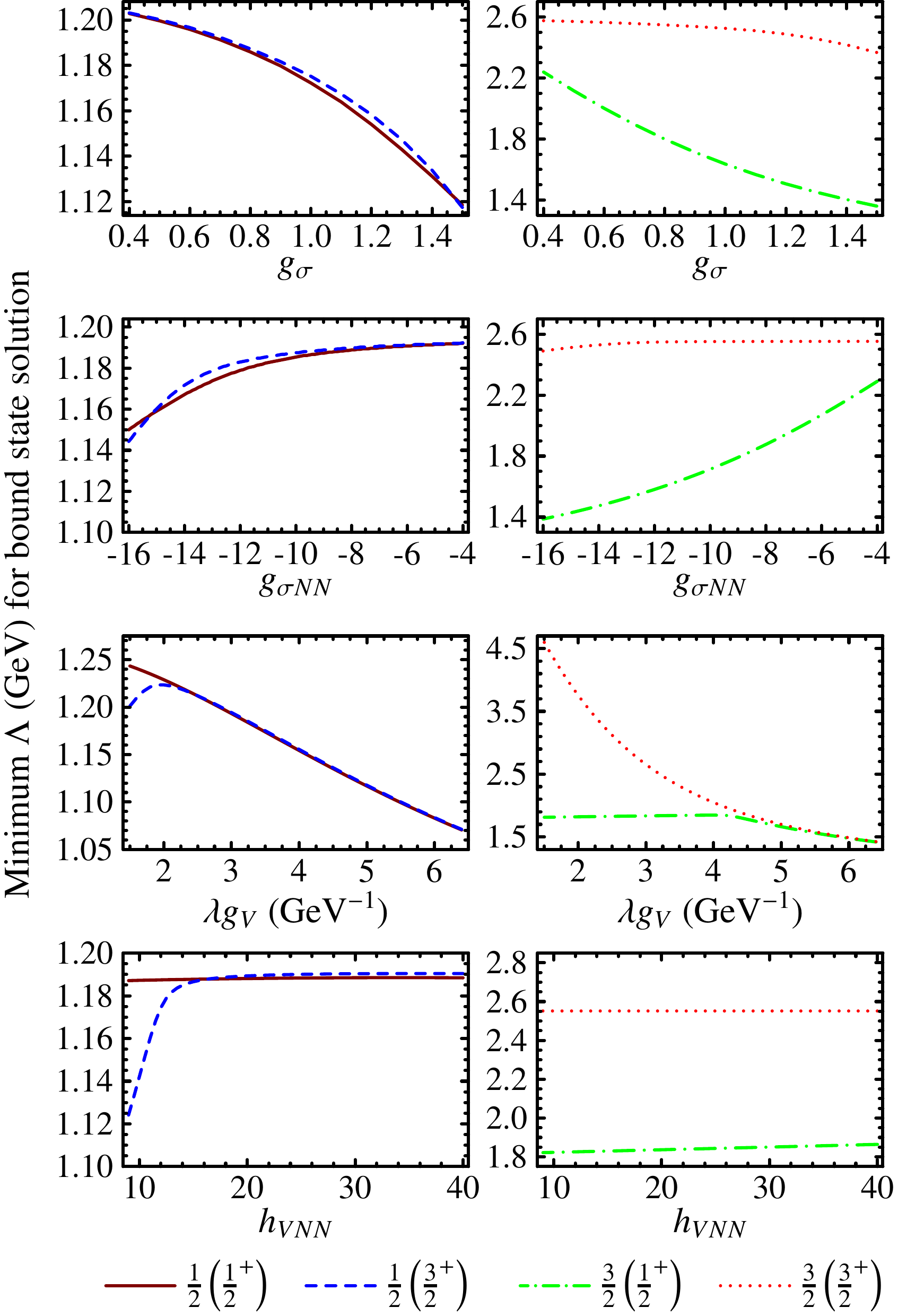}
\caption{Dependence of the minimum cutoff $\Lambda$ on the coupling constants.}
\label{changeconstant}
\end{figure}

In this work, we scan  $g_\sigma$, $\lambda g_V$, $g_{\sigma NN}$, and $f_{VNN}$ in the ranges of $0.5\sim 2.0$ times of the values in Table~\ref{parameters}. Then we search for the minimum cutoff $\Lambda$ allowing for bound-state solutions. The numerical results are presented in Fig.~\ref{changeconstant}.

For the $I(J^P)=\frac{1}{2}(\frac{1}{2}^+)$ and $I(J^P)=\frac{1}{2}(\frac{3}{2}^+)$ $DD^*N$ configurations, we find that the minimum $\Lambda$ is in the ranges of $1.05\sim1.20$ GeV for the existence of bound-state solutions. This implies that the $I=\frac{1}{2}$ $DD^*N$ states are robust hadronic candidates even if the coupling constants $g_\sigma$, $\lambda g_V$, $g_{\sigma NN}$, and $f_{VNN}$ are allowed to vary by 100\% from their central values given in Table~\ref{parameters}.

For the two $I=\frac{3}{2}$ configuration, the  existences of bound state solutions are much more sensitive to the   coupling constants. From the second column of Fig.~\ref{changeconstant}, we can read the following:
\begin{enumerate}
\item If we enlarge the interaction strength of the $\sigma$ exchange, it is possible to find bound-state solutions for $\Lambda<1.4$ GeV in the $I(J^P)=\frac{3}{2}(\frac{1}{2})$ configuration. However, the configuration $I(J^P)=\frac{3}{2}(\frac{3}{2})$ with higher isospin and spin  is much more difficult to form a bound state. In Fig.~\ref{changeconstant}, one can see that bound-state solutions for $I(J^P)=\frac{3}{2}(\frac{3}{2})$  exist for  $\Lambda>2$ GeV when we scan $g_\sigma$ and $g_{\sigma NN}$.
\item We also search for  bound-state solutions for the $I(J^P)=\frac{3}{2}(\frac{1}{2}^+)$ and $I(J^P)=\frac{3}{2}(\frac{3}{2}^+)$ $DD^*N$ systems when we vary $\lambda g_V$ in the range of 0.5 to 2.0 times the original value. For the $I(J^P)=\frac{3}{2}(\frac{1}{2}^+)$ configuration, there exist bound-state solutions with a cutoff $\Lambda$ from about 1.8 to 1.5 GeV. For the $I(J^P)=\frac{3}{2}(\frac{3}{2}^+)$ configuration, we find bound-state solution with $\Lambda\approx4.5$ GeV when taking
$\lambda g_V=1.6$ GeV$^{-1}$ (about half of the value in Table~\ref{parameters}). But we find that the minimum $\Lambda$ needed for the existence of loosely bound-state solutions in the $I(J^P)=\frac{3}{2}(\frac{3}{2}^+)$ configuration rapidly decreases to 1.5 GeV when we enlarge  $\lambda g_V$ to 6.2 GeV$^{-1}$ (about half of the value in Table~\ref{parameters}).
\item If we vary $h_{VNN}$ from $10$ to $40$, the minimum cutoff $\Lambda$ needed for the existence of bound-state solutions is not sensitive to $h_{VNN}$. For the $I(J^P)=\frac{3}{2}(\frac{1}{2})$ and $I(J^P)=\frac{3}{2}(\frac{3}{2})$ configurations, bound states exist when the cutoff $\Lambda$ reaches  1.8 and 2.56 GeV, respectively.
\end{enumerate}

In our scheme, the qualitative conclusions about the $DD^*N$ systems with $I(J^P)=\frac{1}{2}(\frac{1}{2}^+)$ and $I(J^P)=\frac{1}{2}(\frac{3}{2}^+)$ are not changed when we vary the coupling constants $g_\sigma$, $g_{\sigma NN}$, $\lambda g_V$, and $h_{vNN}$. However, the existence of good $I(J^P)=\frac{3}{2}(\frac{1}{2}^+)$ and $I(J^P)=\frac{3}{2}(\frac{3}{2}^+)$ $DD^*N$ molecular candidates is highly dependent on the coupling constants.

\section{Summary}\label{summary}

The observations of $T_{cc}$ \cite{LHCb:2021vvq,LHCb:2021auc}, $\Lambda_c(2940)$ \cite{Aubert:2006sp}, and $\Sigma_c(2800)$ \cite{Belle:2004zjl} have provided us a valuable opportunity to deduce the interactions of $DD^*$, $D^*N$, and $DN$, which makes possible the exploration of the double-charm heptaquark states composed of two charmed mesons and one nucleon.

The present work is dedicated to the study of these kinds of double-charm heptaquark states. Based on the deduced $D$-$D^{(*)}$ and $D^{(*)}$-$N$ effective potentials, we adopted the Gaussian expansion method to solve
the three-body Schr\"odinger equations of the $DD^*N$ system. We searched for bound-state solutions of the  $DD^*N$ system, with both $S$-$D$ mixing and coupled-channel effects considered. Our results imply that the $DD^*N$ bound states with $I(J^P)=\frac{1}{2}(\frac{1}{2}^+)$ and $I(J^P)=\frac{1}{2}(\frac{3}{2}^+)$ are good molecular candidates. On the other hand, the $DD^*N$ systems with $I(J^P)=\frac{3}{2}(\frac{1}{2}^+)$ and $I(J^P)=\frac{3}{2}(\frac{3}{2}^+)$ are difficult to form bound states. The possible decay modes, which can be  searched for these predicted $DD^*N$ molecular states, include: (a) the $T_{cc}p$ channel, (b) the channels of $DDp$ associated with pions and photons, (c) the channel of $\Xi_{cc}$ with pions, and (d) the channel of a charmed baryon with a charmed meson.

In the past years, the LHCb Collaboration observed many heavy flavor hadronic states including the $P_c$ states~\cite{LHCb:2015yax,LHCb:2019kea}, $P_{cs}(4459)$~\cite{LHCb:2020jpq}, $X_{0,1}(2900)$~\cite{LHCb:2020bls,LHCb:2020pxc}, and $X(6900)$~\cite{LHCb:2020bwg}. There is no doubt that the LHCb Collaboration has potential in searching for double-charm heptaquark states predicted in this work, especially
with the running of the high-luminosity LHC.

In addition to the $DD^*N$ systems dedicated in this work, we also noticed that some theoretical groups investigated the $\bar{D}^*\bar{D}^{(*)}N$, $B^*B^{(*)}N$, $\bar{D}^{(*)}NN$, and $B^{(*)}NN$ systems \cite{Garcilazo:2017ifi,Yasui:2009bz,Yamaguchi:2013hsa}, which have different quark components from the discussed $DD^*N$ system in this work. Obviously, the present work associated with these studies \cite{Garcilazo:2017ifi,Yasui:2009bz,Yamaguchi:2013hsa} may reflect the aspect of the three-body hadronic molecular states composed of nucleon and charmed/bottom meson. In future, exploring three-body hadronic molecular states will still be an interesting research topic.

\section*{Acknowledgments}

This work is supported by the National Natural Science Foundation of China under Grants No.11735003, No.11975041, No.11961141004, and the fundamental Research Funds for the Central Universities. X.L. would like to thank the  support from the China National Funds for Distinguished Young Scientists under Grant No. 11825503, National Key Research and Development Program of China under Contract No. 2020YFA0406400, the 111 Project under Grant No. B20063, and the National Natural Science Foundation of China under Grant No. 12047501.

\end{document}